\documentclass[a4paper,11pt]{article}
\usepackage[dvips]{graphicx}
\usepackage{cite}

\begin{document}

\title{Aperture optical antennas}

\author{J\'{e}r\^{o}me Wenger}

\date{}

\maketitle

CNRS, Aix-Marseille Universit\'e, Centrale Marseille, Institut Fresnel, UMR 7249, 13013 Marseille, France

Corresponding author: jerome.wenger@fresnel.fr

\vspace{1.2cm}

\section{Introduction}

Light passing in a small aperture is the subject of intense
scientific interest since the very first introduction of the
concept of \textit{diffraction} by Grimaldi in 1665
\cite{Grimaldi65}. This interest is directly sustained by two
facts: an aperture in an opaque screen is probably the simplest
optical element, and its interaction with electromagnetic
radiation leads to a wide range of physical phenomena. As the
fundamental comprehension of electromagnetism as well as the
fabrication techniques evolved during the twentieth century, the
interest turned towards apertures of subwavelength dimensions.
Bethe gave the first theory of diffraction by an idealized
subwavelength aperture in a thin perfect metal layer
\cite{Bethe44}, predicting extremely small transmitted powers as
the aperture diameter decreased far below the radiation
wavelength. These predictions were refuted by the observation of
the so-called extraordinary optical transmission phenomenon by
Ebbesen and coworkers in 1998 \cite{Ebbesen98}, which in turn
stimulated much fundamental research and technology development
around subwavelength apertures and nano-optics over the last
decade \cite{BarnesNat,NovotnyBook,BrongersmaRev}. It is not the
aim of this chapter to review the transmission of light through
subwavelength apertures. Comprehensive reviews can be found in
\cite{ReviewHoles10} and \cite{Genet07}. Instead, this chapter
will focus on subwavelength apertures to reversibly convert freely
propagating optical radiation into localized energy, and tailor
light-matter interaction at the nanoscale. This goes within the
rapidly growing field of optical antennas
\cite{NovotnyRev11,NovotnyRev}, which forms the core of this book.

From a general perspective as discussed in antennas textbooks
\cite{Balanis,Stutzman}, antennas can be classified into four
basic types: electrically small antennas (of very short dimensions
relative to the wavelength), resonant antennas (which include
common designs such as dipole, patch and Yagi-Uda antennas),
broadband antennas (which operate over an wide frequency range,
such as spiral or log-periodic antennas), and lastly aperture
antennas. Apertures thus define a type of antennas on their own,
the aperture opening determining an obvious effective surface for
collecting and emitting waves. The microphone, the pupil of the
human eye and the parabolic reflector for satellite broadcast
reception can all be considered as examples of aperture antennas.
Electromagnetic aperture antennas operate generally at microwave
frequencies, and are most common for space and aircraft
applications, where they can be conveniently integrated into the
spacecraft or aircraft surface without affecting its aerodynamic
profile.

The aim of this chapter is to review the studies on subwavelength
aperture antennas in the optical regime, paying attention to both
the fundamental investigations and the applications.
Section~\ref{Sec:FundAper} reports on the enhancement of
light-matter interaction using three main types of aperture
antennas: single subwavelength aperture, single aperture
surrounded by shallow surface corrugations, and subwavelength
aperture arrays. A large fraction of nanoaperture applications is
devoted to the field of biophotonics to improve molecular sensing,
which are reviewed in Section~\ref{Sec:BioAper}. Lastly, the
applications towards nano-optics (sources, detectors and filters)
are discussed in Section~\ref{Sec:PhotAper}.

\section{Enhanced light-matter interaction on nanoaperture antennas} \label{Sec:FundAper}

\subsection{Single apertures}

\begin{figure*}[t]
\begin{center}
\includegraphics{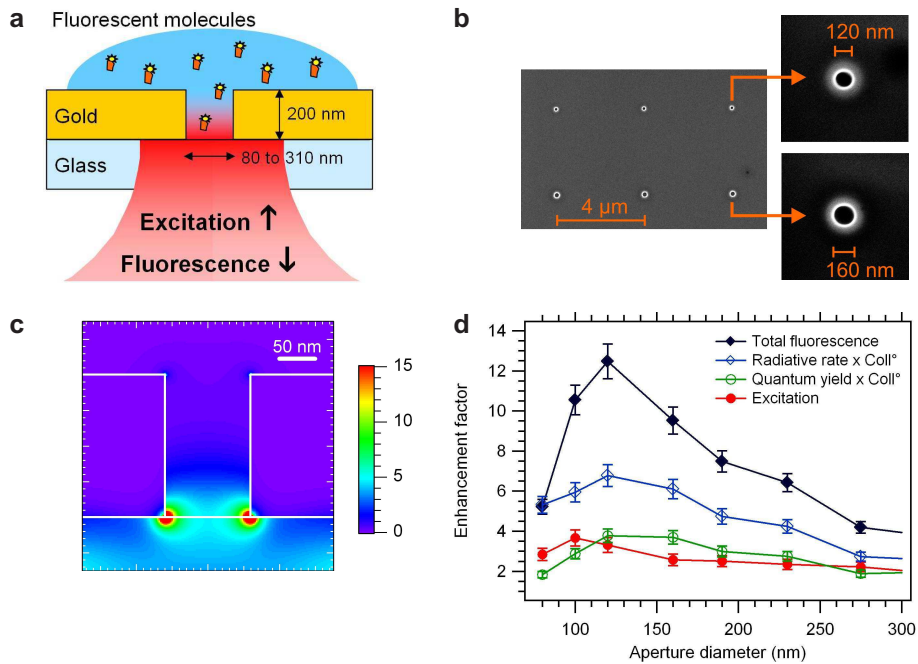}
\caption{(a) Single subwavelength aperture to enhance the
fluorescence emission of molecules located inside the structure
\cite{Wenger08}. (b) Electron microscope images of 120 and 160~nm
apertures milled in gold. (c) Field intensity distribution on a
120~nm water-filled gold aperture illuminated at 633~nm
\cite{AouaniACS10}. (d) Fluorescence enhancement factor and
contributions to nanoaperture enhanced fluorescence of emission
and excitation enhancement, plotted versus the aperture diameter
and normalized to the open solution case, from \cite{Wenger08}.
Figures reproduced with permission: (a,d) $\copyright$ OSA 2008,
(b,c) $\copyright$ ACS 2010.}\label{Fig:SingleAper}
\end{center}
\end{figure*}

The introduction of the concept of subwavelength aperture antennas
to improve optical systems can be attributed to E. H. Synge for
his pioneering vision of scanning near field microscopy
\cite{Synge}. However, the first practical use of subwavelength
apertures to enhance light-matter interaction dates back to 1986
\cite{Fischer86}. In this study, apertures of diameters down to
180~nm fabricated in silver or gold films on glass slides were
used as substrates to detect fluorescent molecules, and clear
indications of fluorescence enhancement were reported.
Fluorescence enhancement for single molecules in a single
subwavelength aperture was reconsidered in 2005, and a 6.5 fold
enhancement of the fluorescence rate per rhodamine 6G molecule was
reported while using a single 150~nm diameter aperture milled in
an opaque aluminum film \cite{Rigneault05}. This result, and the
broad interest devoted to the phenomenon of extraordinary optical
transmission \cite{ReviewHoles10}, led to a large number of
studies to understand the physical origins of the phenomenon,
investigate the role of several design parameters (aperture shape
and dimensions, metal permittivity, metal adhesion layer), and
develop practical applications (Figure~\ref{Fig:SingleAper}).

Influence of the metal layer and aperture diameter was thoroughly
investigated in reference \cite{Davy08}. Comparison with numerical
simulations reveals that the fluorescence enhancement is maximum
when the aperture diameter corresponds to a minimum of the group
velocity of light inside the hole \cite{PopovJOSA}. This provides
a guideline for the design of optimized nanostructures for
enhanced fluorescence detection. For applications in the UV part
of the spectrum, aluminum apertures provide the highest
enhancement factors, with a 20x net increase in tryptophan
molecules fluorescence for 75~nm diameter apertures in aluminum
\cite{BlairUV}. For applications in the near-infrared, gold is the
metal of choice, if sufficient care is taken to properly design
the adhesion layer used between the gold film and glass substrate.
Any increase in the absorption losses due to the adhesion layer
permittivity or thickness was demonstrated to lower the
fluorescence enhancement in subwavelength apertures
\cite{aouaniACS09}, and more generally plasmonic antennas. This
effect was related to a damping of the energy coupling at the
nanoaperture while using absorbant adhesion layers such as
chromium or titanium. Optimisation of the various design
parameters (200~nm thick gold layer, 10~nm titanium dioxide
adhesion layer, 120~nm circular aperture diameter) led to the
largest fluorescence enhancement factor found for single apertures
(25x for Alexa Fluor 647 molecules of 30\% quantum yield in water
solution) \cite{aouaniACS09}. Selecting a molecule with lower
quantum yield would further increase the apparent fluorescence
enhancement factor, with an upper limit of 50x enhancement for
quantum emitters with quantum yield below 1\%
\cite{WengerProcSPIE}. Higher enhancement factors could be in
principle achieved with silver films thanks to lower ohmic losses
in silver as compared to gold. However, the chemical reactivity of
silver makes challenging any experiment with organic fluorophores.

The physical phenomena leading to the fluorescence enhancement in
single subwavelength apertures were investigated in reference
\cite{Wenger08}. By combining methods of fluorescence correlation
spectroscopy and fluorescence lifetime measurements, the
respective contributions of excitation and emission were
quantified (Figure~\ref{Fig:SingleAper}d). Excitation  and
emission enhancement mechanisms were also investigated numerically
\cite{Blair07}, including a spectral study for individual gold
apertures. Fluorescence quenching was clearly observed for
aperture diameters much below the cut-off diameter of the
fundamental mode that may propagate through the aperture. This
explains the existence of an optimum diameter for maximum
enhancement. Lastly, the excitation intensity enhancement was
further confirmed by an independent study monitoring the transient
emission dynamics of colloidal quantum dots in subwavelength
apertures \cite{AouaniACS10}.

Apart form fluorescence, subwavelength apertures were also
demonstrated to enhance a broad range of different light-matter
interactions. Second harmonic generation (SHG) was first
investigated for large ($> 500$~nm) apertures \cite{Dragnea}, then
for subwavelength apertures (circular and triangular) with sizes
down to 125~nm \cite{SHGnanotrou}. The SHG enhancement originates
from a combinaison of field enhancements at the nanoaperture edge
together with phase retardation effects. Triangular nanoapertures
exhibit superior SHG enhancement compared to circular ones, as
expected from their noncentrosymmetric shape. Surface enhanced
Raman scattering (SERS) was also characterized for single
nanoapertures in gold using a non-resonant analyte molecule
\cite{Djaker10}. Thanks to their insensitivity to quenching
losses, SERS and SHG provide essential complementary information
to fluorescence-based studies, specially to quantify the
excitation intensity enhancement at the aperture edge. For
instance, a peak SERS enhancement factor of $2 \times 10^5$ was
quantified for a 100~nm diameter aperture, corresponding to a peak
intensity enhancement $\mathrm{\mid E_{max}\mid^2/\mid E_0 \mid^2
> 200}$ at the aperture edge (for the direction along the incident
polarization). The increase of the local excitation intensity
within subwavelength apertures also leads to other locally
enhanced light-matter interactions, such as erbium up-conversion
luminescence \cite{VerhagenOE09}, or biexciton state formation
rate in semiconductor quantum dots \cite{AouaniACS10}.

The first studies on aperture-enhanced fluorescence were performed
with circular holes, as this shape is polarization insensitive and
relatively simple to fabricate with ion beam milling. Since 2005,
several different aperture shapes have been considered. Slits
\cite{BrongersmaOL09,BrongersmaJPCC10}, rectangles
\cite{Wenger05}, and triangles \cite{NaderTriang} are polarization
sensitive, providing an extra degree of freedom to tune the
electromagnetic distribution inside the aperture, or polarize the
emitted light. Coaxial apertures \cite{Baida06} or ring cavities
\cite{Polman10} display narrower resonances and smaller mode
volumes as compared to circular shapes, suggesting that high
Purcell factors ($>2000$) should be reached with such designs
\cite{Polman10}.

\subsection{Single apertures surrounded by surface corrugations}

Due to its subwavelength dimension, an isolated nanoaperture
antenna does not provide a strong directional control on the light
emitted from the aperture \cite{Degiron04,Davy08}, although edge
effects from the metallic walls have been reported in the case of
single molecule fluorescence experiments \cite{Gersen00}. From
classical antenna theory \cite{Balanis,Stutzman}, the IEEE
directivity $D$ of an aperture antenna can be expressed as $D=4
\pi (area)/\lambda^2$, where $area$ is the effective aperture area
and $\lambda$ is the radiation wavelength. Thus for a circular
aperture of diameter $d$, the directivity is $D = (\pi d /
\lambda)^2$, which shows that the directivity vanishes for a
subwavelength aperture ($d \ll \lambda$).

\begin{figure*}[t]
\begin{center}
\includegraphics{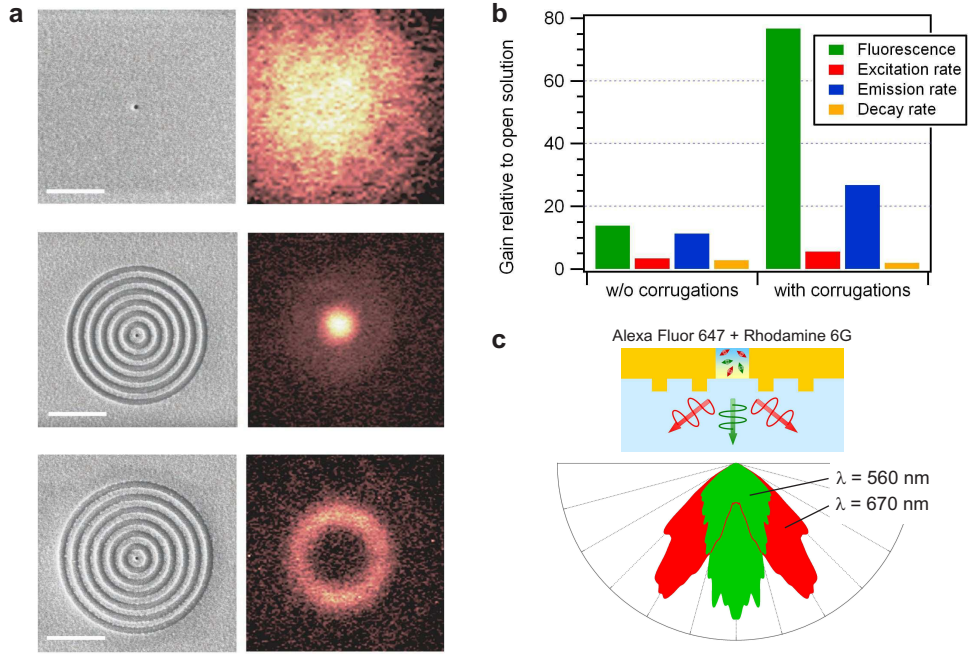}
\caption{(a) Scanning electron microscope images of corrugated
apertures (scale bar 2~$\mu$m) and scanning confocal images of
quantum dot photoluminescence taken in a plane 10~$\mu$m below the
aperture surface (scan size 15~$\mu$m). From top to bottom: single
120~nm aperture, antenna with concentric grooves of 350~nm period,
and antenna with a larger groove period of 420~nm \cite{Jun11}.
(b) Fluorescence enhancement factor and contributions of
excitation and emission gains, in the case of a single aperture
with five corrugations \cite{AouaniNL}. Decay rate corresponds to
the reduction of the fluorescence lifetime. (c) Fluorescence
radiation pattern at two different emission wavelengths
illustrating the directional photon sorting capability of
corrugated apertures, from \cite{AouaniNL2}. Figures reproduced
with permission: (a) $\copyright$ NPG 2011, (b,c) $\copyright$ ACS
2011.}\label{Fig:CorrugAper}
\end{center}
\end{figure*}

Adding concentric surface corrugations (or grooves) on the metal
around the central aperture is an elegant way to increase the
effective aperture area while keeping the subwavelength dimensions
of the aperture \cite{Lezec03} (Figure~\ref{Fig:CorrugAper}a).
This antenna design merges the light localization from the
nanoaperture with the extended near to far-field conversion
capabilities from the concentric grooves. When the corrugations
are milled on the input surface (`reception' mode), the grating
formed by the corrugations provide the supplementary momentum
required to match the incoming light to surface plasmon modes,
which further increase the light intensity at the central
aperture. When the corrugations are milled on the output surface
(`emission' mode), the reverse phenomenon appears, the surface
corrugations couple the surface waves back to radiated light into
the far-field. As the coupling of far-field radiation into surface
plasmon modes is governed by geometrical momentum selection rules,
the coupling occurs preferentially at certain angles for certain
wavelengths. These principles were originally demonstrated in
pioneering transmission experiments on corrugated apertures
\cite{Lezec03}, and confirmed by surface second harmonic
generation experiments \cite{Nahata03}.

Corrugated aperture antennas appear thus as an excellent design to
fully control the radiation from single quantum emitters,
providing high local intensity enhancement together with emission
directionality. Moreover, this design is suitable for the
detection of emitters in liquid solution diffusing inside the
central aperture, thanks to strong localization of light inside
the aperture. Two independent studies have recently demonstrated
these principles for organic fluorescent molecules
\cite{AouaniNL,AouaniNL2} and colloidal quantum dots \cite{Jun11}.
Fluorescence enhancement factors up to 120 fold simultaneous with
narrow radiation pattern into a cone of $\pm15^{\circ}$ have been
reported using a nanoaperture surrounded by 5 circular grooves
\cite{AouaniNL} (Figure~\ref{Fig:CorrugAper}b). The fluorescence
beaming results from an interference phenomenon between the
fluorescence emitted directly from the central aperture and the
surface-coupled fluorescence scattered by the corrugations
\cite{AouaniNL2,Jun11}. Tuning the corrugations period or the
distance from first corrugation to central aperture enables a wide
control over the fluorescence directionality, in very close
fashion to enhanced transmission experiments
\cite{MartinMoreno03,OussamaOE11} (Figure~\ref{Fig:CorrugAper}a
and c). In this framework, the exhaustive investigation of the
design parameter space for enhanced transmission through
corrugated apertures \cite{OussamaOE10} is of major importance to
further optimize the performances of corrugated aperture antennas.
For fluorescence emission, the influence of the number of
corrugations has been quantitatively investigated in
\cite{AouaniOE11}, showing that a single concentric groove already
provides a supplementary 3.5-fold increase in the fluorescence
enhancement as compared to a bare nanoaperture, as suggested
theoretically in \cite{bonod08}. The ability of surface
corrugations to provide for large intensity and radiation
directionality has also stimulated several other studies to
locally enhance Raman scattering \cite{Gordon08} and four wave
mixing \cite{Capasso10}, and to improve the performance of
dipolar-like optical nanoantennas \cite{Crozier11,Crozier11b}.

\subsection{Aperture arrays}

\begin{figure*}[h!]
\begin{center}
\includegraphics{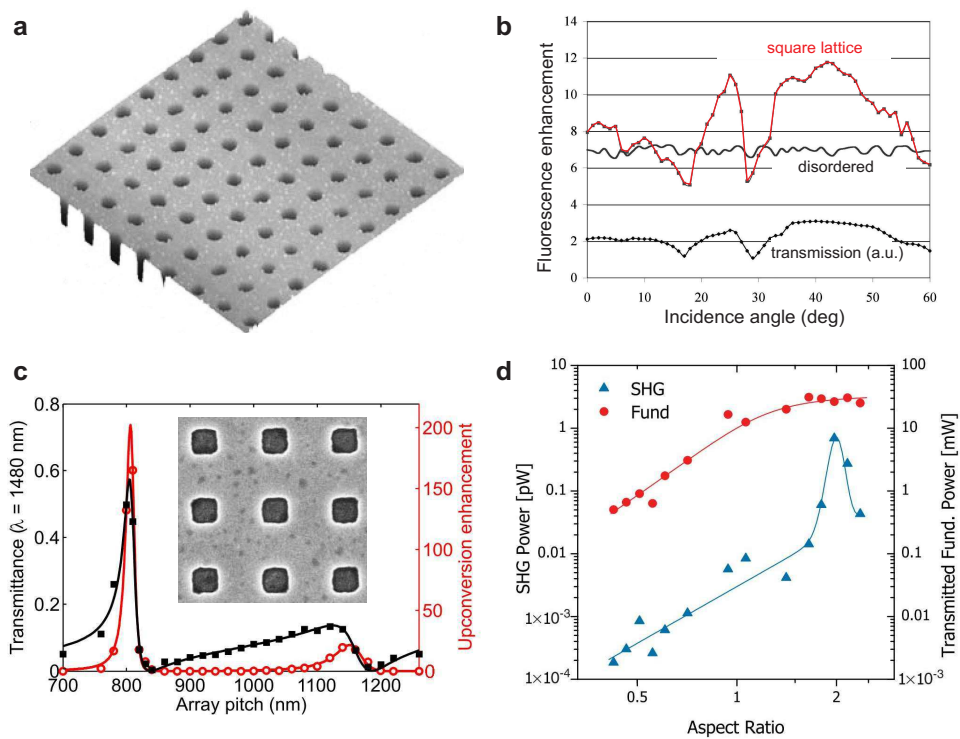}
\caption{(a) AFM images of 200~nm diameter aperture array with
1~$\mu$m period \cite{BlairNano}. (b) Fluorescence enhancement
from Cyanine-5 from a periodic arrangement of 200~nm diameter
apertures in 70~nm thick gold film, with 1~$\mu$m spacing.
Fluorescence from a disordered array and transmission of the
excitation light are also plotted  for reference. Enhancement
factors are normalized to a quartz slide with the same molecular
monolayer, and corrected for fill fraction, adapted from
\cite{BlairNano}. (c) Comparison between the 980~nm erbium
up-conversion enhancement (red) and the transmittance at 1480~nm
(black) as a function of the array period \cite{PolmanOE09}. (d)
Second harmonic generated power (triangles) and fundamental light
transmission (circles) as a function of aperture aspect ratio
\cite{KuipersSHG}. Figures reproduced with permission: (a,b)
$\copyright$ IOP 2004, (c) $\copyright$ OSA 2009, (d) $\copyright$
APS 2006.}\label{Fig:ArrayAper}
\end{center}
\end{figure*}

Arranging the apertures in an array with periodic lattice is
another way to provide for the momentum needed to match the
far-field radiation with surface electromagnetic waves
(Figure~\ref{Fig:ArrayAper}a). These extra coupling capabilities
have largely stimulated several studies on extraordinary optical
transmission for aperture arrays \cite{Ebbesen98,ReviewHoles10}.
Broadly speaking, two types of resonant phenomenon contribute to
explain the transmission peaks observed in far-field and the
intensity enhancement in the near field. The first phenomenon
relies on the resonant excitation of surface plasmon waves at the
metal-dielectric interface, which is obtained at specific incident
angles and wavelengths according to grating diffraction rule. The
second contribution comes from localized plasmon modes on properly
shaped apertures. Combinaison of these two resonant phenomena are
of major interest to locally enhance light-matter interaction, and
control the radiation spectrum, direction and polarization.

Fluorescence enhancement for emitters dispatched over a
subwavelength aperture array was first reported in
\cite{BlairOL,BlairOE,BlairIEEE,BroloJACS} for organic molecules,
then in \cite{BroloJPCB} for colloidal quantum dots. Under
resonant transmission conditions, the fluorescence enhancement
normalized to the aperture array area was estimated to nearly 40
\cite{BlairOL}, while disordered ensemble of apertures lacking
spatial coherence displayed much lower enhancement factors of
about 7 \cite{BlairIEEE} (Figure~\ref{Fig:ArrayAper}b). Maximum
fluorescence signal is found under conditions of enhanced
transmission of the excitation light and excitation of surface
plasmons. Resonant coupling conditions are achieved either by
selecting the incidence angle \cite{BlairOL,BlairIEEE}, or by
adjusting the array lattice \cite{BroloJACS,BroloJPCB,Moyer07}.
Most experiments are performed in transmission mode, yet
reflection mode also displays fluorescence enhancement and beaming
\cite{Zhu10}.

Tuning the aperture shape provides further control on the local
intensity enhancement inside the aperture, as the local resonances
inside the aperture are independent on the incident angle.
Enhancement of erbium ions photoluminescence and up-conversion
luminescence was demonstrated for arrays of annular apertures,
which exhibit a strong transmission resonance
\cite{PolmanAPL09,PolmanOE09} (Figure~\ref{Fig:ArrayAper}c).
Changing the aperture shape also influences the amount of second
harmonic generated by the metallic aperture arrays
\cite{BlairSHG,KuipersSHG}. For rectangular apertures the maximum
second harmonic enhancement is obtained for the shape
corresponding to the cutoff (or equivalently slow propagation) of
the fundamental wavelength through the apertures \cite{KuipersSHG}
(Figure~\ref{Fig:ArrayAper}d) . A similar effect was observed for
fluorescence on single apertures \cite{Wenger05,Davy08}.

\section{Biophotonic applications of nanoaperture antennas} \label{Sec:BioAper}

\subsection{Enhanced fluorescence detection and analysis}

The confinement of light within a single subwavelength aperture
and the local electromagnetic intensity increase are of major
interest to develop new methods for fluorescence analysis down to
the single emitter level. This subsection describes the different
approaches along that direction.

\paragraph{Single molecule fluorescence spectroscopy in liquids}

The smallest volumes that can be achieved by diffraction-limited
confocal microscopy are about a fraction of femtoliter (1 fL =
1~$\mu$m$^3$). To ensure that only one molecule is present in such
volumes, the concentration has to be lower than 10 nanomolar.
Unfortunately, this concentration is too low to ensure relevant
reaction kinetics and biochemical stability, which typically
require concentrations in the $\mu$M to the mM range
\cite{Levene03,Samiee05,CraigheadReview2}. There is thus a very
large demand for nanophotonic structures to overcome the limits
set by diffraction, in order to (i) enhance the fluorescence
brightness per emitter, and (ii) increase the range of available
concentrations by reducing the observation volume. Several
photonic methods have been developed during the last decade, as
reviewed in \cite{WengerIJMS}. Among them, subwavelength apertures
bear the appealing properties of providing the smallest volumes
and the highest fluorescence enhancement to date.

\begin{figure*}[h!]
\begin{center}
\includegraphics[width=12cm]{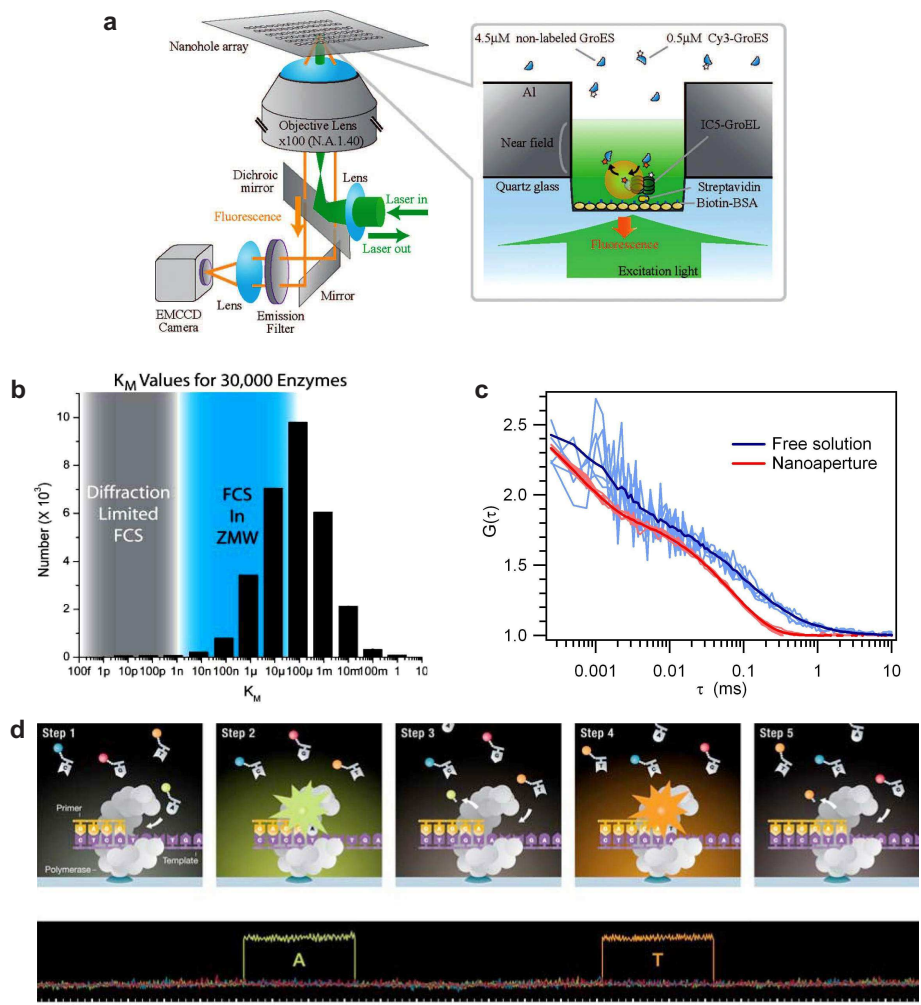}
\caption{(a) Subwavelength aperture antennas for the detection and
analysis of protein-protein interaction \cite{Miyake08}. (b)
Histogram of Michaelis constants for 30,000 enzymes, showing the
range accessible to conventional diffraction-limited FCS and FCS
with nanoapertures (ZMW) \cite{Samiee05}. (c) Fluorescence
correlation functions for 1~s integration time (thin lines). Thick
lines correspond to averaging over 200~s \cite{WengerAC09}. Fast
FCS measurements are enabled by the fluorescence enhancement in a
nanoaperture. (d) Single-molecule real-time DNA sequencing
performed while incorporation of individual nucleotides is
followed, the lower trace displays the temporal evolution of the
fluorescence intensity \cite{Eid09}. Figures reproduced with
permission: (a) $\copyright$ ACS 2008, (b) $\copyright$ BS 2005,
(c) $\copyright$ ACS 2009, (d) $\copyright$ Pacific Biosciences
Inc.}\label{Fig:FCS}
\end{center}
\end{figure*}

The introduction of subwavelength apertures to reduce the analysis
volume in single molecule fluorescence spectroscopy was performed
by the groups of Harold Craighead and Watt Webb in an outstanding
contribution \cite{Levene03}. A subwavelength aperture milled in
an opaque metallic film is an elegant way to generate an analysis
volume much below the diffraction limit (Figure~\ref{Fig:FCS}a),
enabling single molecule analysis at much higher concentrations
(Figure~\ref{Fig:FCS}b). Subwavelength apertures have thus been
termed zero-mode waveguides or ZMW to emphasize the evanescent
nature of the excitation light inside the aperture. A large range
of biological processes have been monitored with single molecule
resolution at micromolar concentrations while using nanoapertures.
This includes DNA polymerase activity \cite{Levene03},
oligomerization of the bacteriophage $\lambda$-repressor protein
\cite{Samiee05}, DNA enzymatic cleavage
\cite{WengerFCCS,WengerAC09}, and protein-protein interactions
\cite{Miyake08}. Moreover, the physical limitation of the
observation volume by the nanoaperture greatly simplifies the
optical alignment for multi-color cross-correlation analysis
\cite{WengerFCCS}. The high fluorescence count rates improve the
signal to noise ratio by over an order of magnitude, enabling a
100-fold reduction of the experiment acquisition time
\cite{WengerAC09} (Figure~\ref{Fig:FCS}c). This offer new
opportunities for probing specific  biochemical reactions that
require fast sampling rates.

\paragraph{DNA sequencing}

The development of personalized quantitative genomics requires
novel methods of DNA sequencing that meet the key requirements of
high-throughput, high-accuracy and low operating costs
simultaneously. To meet this goal, subwavelength apertures are
currently being used as nano-observation chambers for
single-molecule, real-time DNA sequencing \cite{Eid09,Meller10}
(Figure~\ref{Fig:FCS}d). Within each aperture, a single DNA
polymerase enzyme is attached to the bottom surface
\cite{Korlach08}, while distinguishable fluorescent labeled
nucleotides diffuse into the reaction solution. The sequencing
method records the temporal order of the enzymatic incorporation
of the fluorescent nucleotides into a growing DNA strand
replicate. Each nucleotide replication event last a few
millisecond, and can be observed in real-time. Currently, over
3000 nanoaperture are operated simultaneously, allowing
straightforward massive parallelization \cite{Eid09}.

\paragraph{Live cell membrane investigations}

\begin{figure*}[h!]
\begin{center}
\includegraphics{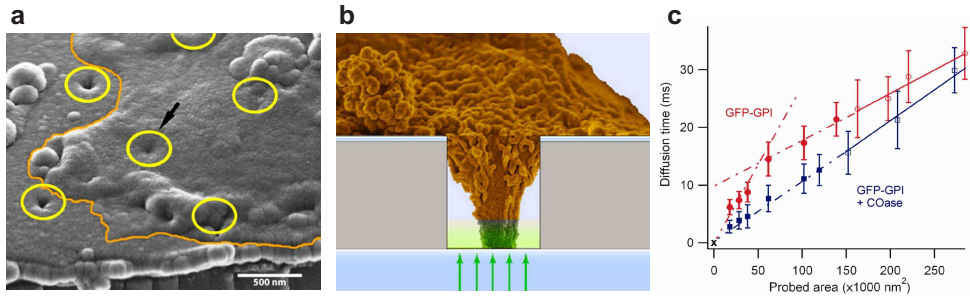}
\caption{(a) Tilted scanning electron microscope view of
cross-sectional cuts of nanoapertures. Cell membranes have been
outlined (orange line), and aperture locations have been circled
(yellow). Cell membrane spanning a nanoaperture dips down (arrow),
suggesting membrane invagination \cite{CraigMemb07}. (b)
Cross-sectional cartoon of cell invaginating into a subwavelength
aperture (not drawn to scale; the shape of the membranous
extension into the aperture is hypothetical) \cite{CraigMemb07}.
(c) Molecular diffusion times versus aperture area for untreated
GFP-GPI protein and GFP-GPI with 1~U/mL cholesterol oxidase
(COase) to reveal for transient diffusion regimes related to
membrane heterogeneities on the nanometer scale
\cite{WengerMemb2}. Figures reproduced with permission: (a,b)
$\copyright$ IOP 2007, (c) $\copyright$ BS 2007.}\label{Fig:Memb}
\end{center}
\end{figure*}

Investigating the cell membrane organization with nanometer
resolution is a challenging task, as standard optical microscopy
does not provide enough spatial resolution while electron
microscopy lacks temporal dynamics and cannot be easily applied to
live cells \cite{MarguetEMBO}. A subwavelength aperture provides a
promising tool to improve the spatial resolution of optical
microscopy. Contrarily to near-field scanning optical microscopy
(NSOM), the subwavelength aperture probe is fixed to the
substrate, with a cell being attached above
(Figure~\ref{Fig:Memb}a,b). The aperture works as a pinhole
directly located under the cell to restrict the illumination area.
Diffusion of fluorescent markers incorporated into the cell
membrane provide the dynamic signal, which is analyzed by
correlation spectroscopy to extract information about the membrane
organization \cite{CraigMemb,CraigMemb2}. To gain more insight
about the membrane organization, measurements can be performed
with increasing aperture diameters
\cite{WengerMemb2,LaureW,WengerMemb} (Figure~\ref{Fig:Memb}c).
This set of experiments demonstrated the aperture limited the
observed membrane area, and did not significantly alter the
diffusion process within the membrane. It was also shown that
fluorescent chimeric ganglioside proteins partition into
structures of 30~nm radius inside the cell membrane
\cite{WengerMemb2}. The combinaison of nanoapertures with
fluorescence correlation spectroscopy on membranes provide a
method having both high spatial and temporal resolution together
with a direct statistical analysis. The major limitation of this
method is directly related to the need for cell membranes to
adhere to the substrate. Cell membrane invagination within the
aperture was shown to depend on the membrane lipidic composition
\cite{CraigMemb2} and on actin filaments \cite{CraigMemb07}. To
further ease cell adhesion, and avoid membrane invagination
issues, planarized 50~nm diameter apertures have been recently
introduced \cite{CraigMemb11}. The planarization procedure fills
the aperture with fused silica, to achieve no height distinction
between the aperture and the surrounding metal. The technique
provides 1~$\mu$s and 60~nm resolution without requiring
penetration of the membrane into the aperture.

\paragraph{Trapping}

\begin{figure*}[h!]
\begin{center}
\includegraphics{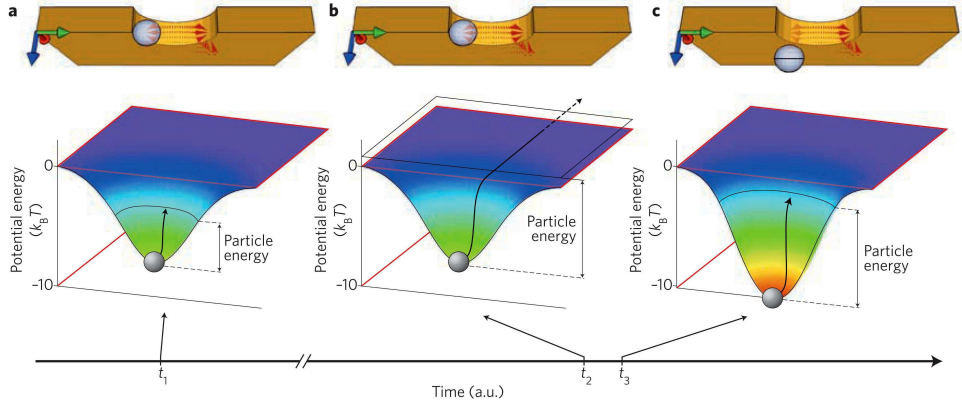}
\caption{Self-induced back-action trapping, adapted from
\cite{QuidantReview}. (a) The particle is localized in the
aperture at time $t_1$  with moderate kinetic energy. (b) During a
high-energy event at time $t_2$, the object may escape the
aperture. (c) As the particle moves out of the aperture at time
$t_3$, the SIBA force increases the potential depth to maintain
the object within the trap. Figures reproduced with permission
$\copyright$ NPG 2011.}\label{Fig:Trap}
\end{center}
\end{figure*}

Optical tweezers have become a powerful tool for manipulating nano
to micrometer sized objects, with applications in both physical
and life sciences. To overcome the limits set by the diffraction
phenomenon in conventional optics and extend optical trapping to
the nanometer scale, metallic nanoantennas have been recently
introduced and reviewed in \cite{QuidantReview}. Most works on
plasmon nano-optical tweezers relie on a strong enhancement of the
local intensity provided by the nanoantenna. This approach induces
high local intensities, often above the objects damage threshold.
A subwavelength aperture can solve this challenge, and achieve
more than an order of magnitude reduction in the local intensity
required for optical trapping \cite{Quidant09}
(Figure~\ref{Fig:Trap}). The optical trapping method is called
self-induced back-action (SIBA), as the trapped object plays an
active role in enhancing the restoring force. Trapping of a single
50~nm polystyrene sphere was demonstrated based on the
transmission resonance of a 310~nm diameter aperture in a gold
film \cite{Quidant09}. Remarkably, the local intensity inside the
aperture is only enhanced by a moderate factor of seven.
Low-intensity optical trapping of nanoparticles enables new
opportunities for isolating and studying biological nano-objects,
such as viruses. This trapping method can also be coupled directly
to sensing and sorting based on transmission changes through the
aperture.

\subsection{Molecular sensing and spectroscopy with aperture arrays}

Sensors able to detect a specific type of molecules in real-time
and with high sensitivity are a subject of intense research, and a
major drive for the field of plasmonics. Compared to other
nanoantenna arrays designed for plasmon-enhanced sensing,
subwavelength apertures bear the specific advantages of presumably
better robustness and higher reproducibility, as the fabrication
is comparatively simpler and the mode of operation does not rely
on ultra-high intensity enhancement. This subsection reviews the
different spectroscopic applications of subwavelength aperture
arrays.

\paragraph{Surface plasmon resonance spectroscopy}

\begin{figure*}[h!]
\begin{center}
\includegraphics{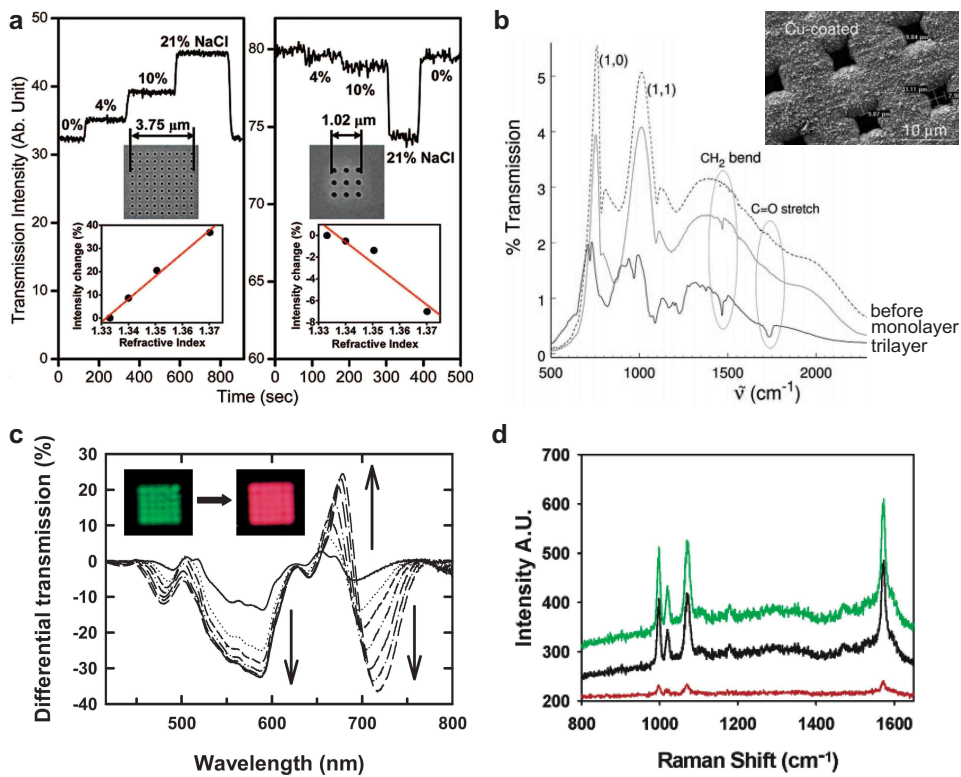}
\caption{(a) Transmission response to surface refractive index
change from a 9x9 and a 3x3 nanohole array \cite{LarsonNL08}. (b)
Infrared transmission spectra of copper-coated mesh before and
after coating with 1-hexadecanethiol. There is significant damping
of the transmission with the successive coatings, molecular
absorptions are indicated with the solid ovals \cite{Coe3}. (c)
Differential  transmission of an array (period 390 nm, diameter
260 nm, depth 180 nm) covered with a spiropyran-doped PMMA film
after different UV irradiation times (1– 130 s). Arrows indicate
the variation for increasing irradiation time. The insets show
transmission images of the array before and after irradiation
\cite{Dintinger1}. (d) Nanoaperture-enhanced Raman spectra of
benzenethiol. The red spectrum was obtained from an unpatterned
portion of the film; the black spectrum was obtained from a
nanoaperture array with 450~nm lattice spacing. The green spectrum
was corrected for the reduced geometric area on the array
\cite{Schatz07}. Figures reproduced with permission: (a)
$\copyright$ ACS 2008, (b) $\copyright$ ACS 2006, (c) $\copyright$
Wiley-VCH 2006, (d) $\copyright$ ACS 2007.}\label{Fig:Spectro}
\end{center}
\end{figure*}

Conventional surface plasmon resonance (SPR) sensing is based on
the excitation of extended surface plasmon modes on a thin metal
layer through prism coupling in the Kretschmann configuration.
This method has proven to be sensitive to tiny refractive index
changes at the metal surface down to the molecular monolayer
level. The transmission of light through aperture arrays is also
sensitive to refractive index changes around the metal
\cite{Krishnan01} (Figure~\ref{Fig:Spectro}a). Currently, the
sensitivity is comparable to other SPR devices, and molecular
binding events can be followed dynamically by measuring a spectral
shift in the transmitted light \cite{BroloLG,Larson05,Tetz06}.
Nanoaperture arrays appear thus well suited for dense integration
in a sensor chip in a collinear optical arrangement providing a
simpler setup and a smaller probing area than the typical
Kretschmann configuration. Current research directions include
lan-on-chip integration with microfluidic systems
\cite{BroloAC07,Sharpe08,LesuffleurOE08}, increasing the
sensitivity \cite{LesuffleurAPL07,LesuffleurAC09} and multiplexing
the amount of extracted information \cite{LarsonNL08,LarsonAC08}.

Isolated apertures or disordered patterns of apertures in thin
gold films also exhibit a localized surface plasmon resonance
leading to a peak in the extinction spectrum in the near-infrared
region which can be used for sensing applications
\cite{Kall04,KallNL05}. This type of device has been successfully
employed to monitor membrane biorecognition events
\cite{Kall05,Hook07}, and selective sensing for cancer antigens
\cite{Schultz07}. Aperture sensors can also be designed to work as
nanopores, with the liquid flowing across the aperture arrays
\cite{Hook10}. This configuration further improves the uptake rate
of biomolecules and thus the sensing temporal resolution.

\paragraph{Enhanced absorption and fluorescence spectroscopy}

Nanoaperture arrays tuned for resonant transmission in the
infrared were demonstrated to enhance the absorption of molecules
adsorbed on the array by at least two orders of magnitude
\cite{Coe1} (Figure~\ref{Fig:Spectro}b). Enhanced absorption
spectroscopy can thus be used to monitor catalysis process
\cite{Coe2} or phospholipid assembly \cite{Coe3}. The absorption
enhancement is related to a long lifetime of surface plasmon modes
in the infrared, which increases the interaction probability
between molecules and light. Absorption enhancement of electronic
transitions was also reported in the visible \cite{Dintinger1},
with lower enhancement factors of about one order of magnitude
related to shorter plasmon lifetime or increased propagation
losses (Figure~\ref{Fig:Spectro}c). Absorption enhancement is
motivating new time-resolved spectroscopy studies to explore
transient molecule-plasmon states \cite{Dintinger2,Salomon09}.

Enhancement of the fluorescence process was also used to perform
DNA affinity sensing on aperture arrays spotted with probe DNA
sequences \cite{BlairNano}. Performing detection on the back-side
of the aperture sample provides high signal-to-background
rejection, and enables real-time detection. Interestingly, capture
of target molecules can be further improved by UV photoactivation
of the aperture array silanized bottom surface \cite{Blair11}.
This photoactivation procedure is a promising strategy to achieve
localization of target molecules to the region of plasmonic
enhancement.

\paragraph{Surface enhanced Raman spectroscopy}

Metallic nanostructures have attracted much interest over the last
years to realize efficient and reproducible media for
surface-enhanced Raman scattering (SERS) spectroscopy
\cite{Tsuruk08}. The major aim is to develop SERS substrates
combining high sensitivity with control and localization of the
regions leading to high SERS enhancement. Among the different
strategies being explored, subwavelength apertures milled in noble
metal films realize promising substrates thanks to their rational
and tunable design, controlled surface enhancement,
surfactant-free fabrication and intrinsic robustness
(Figure~\ref{Fig:Spectro}d). The first SERS study with
nanoaperture arrays was performed on resonant oxazine~720 dyes
\cite{Brolo04}. The enhancement factor reached a maximum for the
array that presented the largest transmission at the excitation
wavelength of the laser, which was confirmed by several other
studies \cite{Bahns06,Rowlen07,Golden07,Wallace08,Brolo08}.
Reference \cite{Schatz07} presents a remarkable quantitative study
to determine the absolute Raman scattering enhancement factors for
nanoaperture arrays in a silver film as a function of aperture
lattice spacing, and using a nonresonant analyte. Maximum
area-corrected SERS enhancement factor of $6 \times 10^7$ was
obtained, which was attributed to two distinct sources: plasmons
localized near the aperture edges and nanometer scale roughness
associated with the silver film. Even higher enhancement factors
could be reached by optimising further the aperture dimensions
\cite{Djaker10}, or by performing SERS on more complex aperture
antennas arrays, such as double-hole arrays \cite{Lesuffleur07} or
combined aperture-nanoparticle pairs \cite{Xu08}. Lastly, the
reproducibility of the SERS measurements was assessed in
\cite{Bahns09} for 2D hexagonal gold aperture arrays. Overall,
area-averaged deviation from measurement to measurement ranged
from 2 - 15\%, which makes nanoaperture arrays a very competitive
platform for sensitive and reproducible SERS.

\section{Nanophotonic applications of nanoaperture antennas}\label{Sec:PhotAper}

\subsection{Photodetectors and filters}

Probably the most straightforward use of subwavelength aperture
devices for photonic applications employs them as wavelength
filters and polarisers. Periodic arrays display well-defined
resonances depending on the lattice symmetry, period, aperture
shape and lattice symmetry \cite{Ebbesen98,ReviewHoles10,Genet07},
and already an isolated rectangular aperture can be made as a
wavelength and polarization sensitive filter \cite{Degiron04}.
Adding an elliptical plasmonic grating around a central
subwavelength aperture realizes an antenna acting as a miniature
planar wave plate \cite{Drezet08}. The difference between the
short and long axis of each ellipsis introduces a phase shift on
the surface waves enabling the operation as a quarter wave plate.

\begin{figure*}[h!]
\begin{center}
\includegraphics{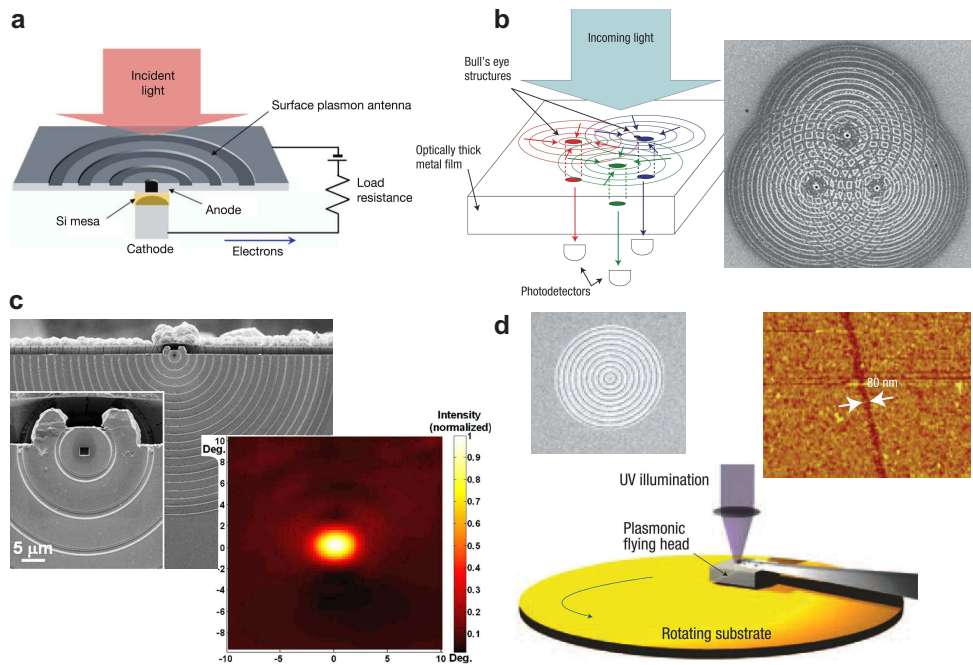}
\caption{(a) Ultrafast nanophotodiode consisting of a 300~nm
silicon photoelectric element integrated into a corrugated
aperture antenna \cite{Ishi05}. (b) Spatial filtering for the
incoming white light through three overlapping corrugated aperture
antennas. The different colours are separated as they couple to
different gratings and are redirected towards three distinct
photodetectors integrated inside the apertures. The inset shows an
experimental realization with grating periods of 730~nm (top
antenna), 630~nm (left), and 530~nm (right) \cite{Laux08}. (c)
Quantum cascade laser integrated with a corrugated aperture
collimator, and measured far-field intensity distribution
\cite{CapassoAPL08}. (d) High-throughput maskless nanolithography
using aperture antennas arrays, and AFM image of a pattern with
80~nm linewidth on the thermal photoresist \cite{Zhang08}. Figures
reproduced with permission: (a) $\copyright$ JJAP 2005, (b)
$\copyright$ NPG 2008, (c) $\copyright$ AIP 2008, (d) $\copyright$
NPG 2008.}\label{Fig:Phot}
\end{center}
\end{figure*}

A major bottleneck in the development of ultrafast photodetectors
can be summarized as follows: to reduce the photodiode capacitance
and increase its operational speed, the active semiconductor
region needs to be reduced to sub-micron dimensions, yet this tiny
active area also leads to low quantum efficiency and low
sensivity. The ability of shallow surface corrugations to
concentrate light to the central aperture \cite{Lezec03,Nahata03}
is highly beneficial to solve this challenge. Periodic
corrugations on the metal surface act as resonant antennas to
capture the incoming light, which can then be concentrated into
one or more apertures filled with photovoltaic elements. Hence
smaller photovoltaic elements can still detect an enlarged amount
of light energy. This principle was first demonstrated with 300~nm
diameter silicon photodiode surrounded by a 10~$\mu$m grating
antenna \cite{Ishi05} (Figure~\ref{Fig:Phot}a), and was recently
extended to telecom wavelengths with germanium photodiode
\cite{Ren11}. Moreover, appropriate texturing of metal surfaces
enables sorting the incoming light according to wavelength  and
polarization, before refocusing the energy into individual
photodetector elements \cite{Laux08} (Figure~\ref{Fig:Phot}b).
This photon-sorting capability provides a new approach for
spectral and polarimetric detectors with highly integrated
architectures.

\subsection{Nanosources}

The antenna capabilities of corrugated apertures have attracted
much attention to improve the performance of vertical-cavity
surface-emitting lasers \cite{Guo07} and quantum cascade lasers
emitting in the infrared \cite{Capasso08,CapassoAPL08}. Surface
plasmons are used to shape the beams of edge or vertical surface
emitting semiconductor lasers and greatly reduce their large
intrinsic beam divergence (Figure~\ref{Fig:Phot}c). Using
concentric semi-circular grating structure, a collimated laser
beam was achieved with remarkably small divergence angles of
2.7$^{\circ}$ and 3.7$^{\circ}$, which correspond to a reduction
by a factor of 30 and 10, compared to those without plasmonic
collimation \cite{CapassoAPL08}. The grating antenna can also be
modified to control the polarization of the laser beam, or achieve
complex wavefront engineering \cite{Capasso08}. As for lasers, the
operation of light-emitting diodes (LEDs) can benefit from
aperture antennas. Aperture arrays engraved in one of the
electrodes provide an outcoupling mechanism for the trapped
electromagnetic energy as well as a control over the emission
properties \cite{Liu05}.

The strong localization of electromagnetic energy with aperture
antennas has stimulated a broad interest for achieving maskless
subwavelength optical lithography, as an alternative to
electron-beam and scanning-probe lithography
(Figure~\ref{Fig:Phot}d). Such direct lithography writing would be
activated directly in the near field of the aperture, which makes
it very difficult to scan the aperture above the surface at high
speed. The first report introduced a self-spacing air bearing to
fly the aperture about 20~nm above the photoresist with spinning
speeds up to 12 m/s \cite{Zhang08}. Recent advances have reported
achievement of patterning with linewidth down to 50~nm and a
patterning speed of 10~mm/s \cite{Hahn09}. The same technique
could also be applied to plasmonic-enhanced data storage, further
improving the blu-ray disc capacity by about 2-fold
\cite{Hahn09b}.

\section{Conclusion}

Compared to nanoparticle-based plasmonic antennas, aperture
antennas bear the essential advantage of providing a high contrast
between the strong opacity of the metallic film and the aperture
element. Although the local field enhancement are not as strong as
in the case of bowtie antennas for instance
\cite{Pohl05,Moerner09,Punj13}, aperture antennas are comparatively
simpler to fabricate and to implement, and readily provide for the
high reproducibility needed in biosensing applications. Texturing
the metal around the apertures opens novel opportunities to
control the antenna operation. Further developments and
applications are thus expected in the years to come in a variety
of areas.

\section*{Acknowledgements}

I am deeply indebted to many at the Fresnel Institute and the
Laboratoire des Nanostructures at the Institut de Science et
d'Ing\'{e}nierie Supramol\'{e}culaires. I would like to gratefully
acknowledge the collaboration with Herv\'{e} Rigneault and Thomas
Ebbesen, together with my coworkers or collaborators: Heykel
Aouani, Steve Blair, Nicolas Bonod, Elo\"{\i}se Devaux,  Davy
G\'{e}rard, Pierre-Fran\c{c}ois Lenne, Oussama Mahboub, Evgeny
Popov, and Brian Stout.


\end{document}